%% file: main.tex
\documentclass{Interspeech2024}

\interspeechcameraready

\usepackage{amsmath}
\usepackage{booktabs}
\usepackage{graphicx}
\usepackage{microtype}
\usepackage{xspace}
\usepackage{url}
\usepackage{tikz}
\usepackage{comment}
\usepackage{multirow}
\usepackage{xcolor}
\usepackage{pifont}
\usepackage[prependcaption,textsize=scriptsize]{todonotes}
\usepackage{tabularx}
\usepackage{hyperref}
\definecolor{indiagreen}{HTML}{138808}%
\definecolor{papaya}{HTML}{EE892F}%
\definecolor{mygreen}{HTML}{008000}%
\definecolor{mypurple}{HTML}{9966CC}%
\definecolor{mygray}{HTML}{696969}

\newcommand{\cmark}{\textcolor{black}{\ding{51}}}
\newcommand{\xmark}{\textcolor{gray}{\ding{55}}}

\newcommand{\mypar}[1]{\vspace{2mm}\noindent\textbf{#1}}
\title{Towards generalisable and calibrated audio deepfake detection with self-supervised representations}

\name[affiliation={1}]{Octavian}{Pascu}
\name[affiliation={1,2}]{Adriana}{Stan}
\name[affiliation={1}]{Dan}{Oneata}
\name[affiliation={3}]{Elisabeta}{Oneata}
\name[affiliation={1}]{Horia}{Cucu}

\address{
    $^1$ \textsc{Politehnica} Bucharest, Romania \\
    $^2$ Technical University of Cluj-Napoca, Romania \\
    $^3$ Bitdefender, Romania}

\email{}

\def\eg{\textit{e.g.}\xspace}
\def\ie{\textit{i.e.}\xspace}
\def\etal{\textit{et al.}\xspace}

\newcommand{\inthewild}{In-the-Wild\xspace} 
\newcommand{\asvspoof}{ASVspoof'19\xspace} 

\newcommand\wavvec{\texttt{wav2vec2}\xspace}
\newcommand\wavlm{\texttt{wavlm}\xspace}

\keywords{%
    Deepfake detection,
    anti-spoofing, 
    pretrained representations,
    generalisation,
    calibration,
    reliability estimation.
}

\begin{document}

\maketitle

\begin{abstract}
Generalisation---the ability of a model to perform well on unseen data---is crucial for building reliable deepfake detectors.
However, recent studies have shown that the current audio deepfake models fall short of this desideratum.
In this work we investigate the potential of pretrained self-supervised representations in building general and calibrated audio deepfake detection models.
We show that large frozen representations coupled with a simple logistic regression classifier are extremely effective in achieving strong generalisation capabilities:
compared to the RawNet2 model,
this approach reduces the equal error rate from 30.9\% to 8.8\% on a benchmark of eight deepfake datasets, while learning less than 2k parameters.
Moreover, the proposed method produces considerably more reliable predictions compared to previous approaches making it more suitable for realistic use.

\end{abstract}

\section{Introduction}
\label{sec:intro}

The ability to synthetically generate audio data is constantly improving~\cite{ liu2023audioldm,masood2023deepfakes,kim23k_interspeech}.
While these advancements have many beneficial applications
(such as allowing speech-impaired persons to recover their voices or creating digital art and entertainment content),
they can also serve malicious purposes (\eg, cloning officials' voices to spread misinformation).
Synthetic speech detection (or audio deepfake detection) attempts to prevent such misuses of the technology
by developing methods that automatically estimate whether a given audio is real (bonafide) or fake (spoofed).
While there is a sustained ongoing effort on this task~\cite{wang2022odyssey,zhang2023icml,rawnet2},
we argue that for the ensuing methods to be effective, they should strive for two properties:
they should generalise and be trustworthy (\ie, well calibrated).

\mypar{Generalisable detection methods.}
Since synthesis methods are continuously evolving, it is unreasonable to expect 
that we will have access to training data similar to that encountered in practice.
\textit{Generalisation} is the capability of a model to perform well on data not seen at training time.
However, Müller \etal ~\cite{muller2022interspeech} have shown that the generalisation abilities of popular audio deepfake detectors have been overestimated.
They evaluate twelve top-performing detection models and
show that none of them are able to generalise on an out-of-distribution dataset.

A possible explanation for the poor generalisation performance is related to the preprocessing peculiarities exhibited by the training dataset (\asvspoof  \cite{WANG2020101114})---%
the silence duration~\cite{muller2021speech} and the bitrate information~\cite{borzi2022synthetic} correlate with the ground truth.
Given that the best deepfake detection models are high-capacity,
they can easily learn such low-level spurious features.
In this paper, we take  an under-explored path and assess the power of strong pretrained representations to improve the generalisation capabilities of audio deepfake detectors.
Our approach is motivated by the strong generalisation results shown in~\cite{ojha2023cvpr}
in the context of image deepfake detection.

While self-supervised representations were also applied to the detection of spoofed speech,
previous work only used smaller representations with modest results \cite{wang2022odyssey, tak2022odyssey}.
Moreover, these approaches did not focus on generalisation and
were tested only %
on the \asvspoof dataset,
and not other challenging datasets, such as \inthewild ~\cite{muller2022interspeech}.
Other works finetuned these features or integrated them in more complex systems \cite{wang2022odyssey, xie2023learning},
but in this way they lose the implicit generalisation power.

\mypar{Well-calibrated detection methods.}
Deepfake detectors can be used to make critical decisions,
so it is crucial to be reliable and trustworthy.
If a detector outputs a fakeness score of 0.7 for a series of inputs,
then we should expect that in 70\% of the cases the input is indeed fake. A classifier with this property is known as \textit{well calibrated}.
Current research in general machine learning addresses this aspect,
but surprisingly little work discusses the calibration of deepfake detectors.
Recent works~\cite{guillaro2023cvpr, salvi2023icassp} have tackled a related problem of estimating the uncertainty (or conversely certainty) in a prediction.
Both papers use a similar method:
first train a deepfake detector,
then train a second classifier (using a frozen representation extracted from the detector) to estimate if the predictions of the first are correct or not.
Here, we investigate if it is feasible to use the more direct method of estimating the uncertainty from the output probabilities of the detector, by computing the entropy over the outputs.

\mypar{Our contributions} are as follows:
(i) We propose a simple yet effective approach for detecting spoofed audio signals.
This approach achieves state-of-the-art results
and produces better calibrated outputs,
suitable for reliability estimation.
(ii) We conduct a systematic study over the impact of various factors including the choice of self-supervised representations, the back-end classifier and the amount of training data.
(iii) We benchmark our approach on eight different datasets including partially-spoofed and multilingual data,
laying the grounds for future research on general deepfake detection.

\section{Methodology}
\label{sec:meth}

\input{tab-datasets}

The task of audio deepfake detection is to predict whether a given audio is real (bonafide) or fake (spoof).
We investigate the approach of
first extracting a pretrained representation using frozen self-supervised models and
then training a binary classifier (logistic regression) on top of these representations. 
Logistic regression estimates probabilities, which we use for the task of uncertainty estimation. 
While these components are not novel, their application to this setting is.
As we will see, this solution offers strong generalisation and calibration performance, as opposed to even more elaborate models.

\mypar{Pretrained representations.} We investigate self-supervised representations stemming from the wav2vec 2.0~\cite{wav2vec2} method.
We have chosen this family of models because 
it has proved strong transfer abilities
\cite{yang2021superb}
and comes in multiple variants,
enabling us to assess the importance of various factors, such as model size or pretraining data.
Wav2vec 2.0 was designed to perform unsupervised pretraining on raw audio data and as a result
learns useful speech representations without the need of phonetic or linguistic annotations.
It uses contrastive predictive coding \cite{oord2018cpc} to capture high-level feature and contextual information. 
Several wav2vec 2.0 extensions were subsequently developed.
The first class of models is XLS-R~\cite{xlsr},
which learns cross-lingual speech representations.
This is achieved by adding a shared quantisation module over the feature encoder representations, producing multilingual speech units, and thus sharing acoustic representations across languages.
This model comes in different sizes: 300M, 1B or 2B parameters.
The second class of models is WavLM~\cite{wavlm},
which considered the task of speech denoising in addition to the masked audio prediction task in wav2vec.
The model aims to produce high-level features targeted towards other non-ASR tasks. 
We use all these variants in our experiments.

\mypar{Calibration and reliability estimation.} A classifier is calibrated if its predictions match the accuracy obtained for that particular level of confidence.
We apply the logistic regression classifier which uses the cross-entropy loss.
The cross-entropy loss is a proper loss \cite{blasiok2023neurips}, which improves the calibration properties.
This choice avoids the need of other post-processing techniques such as Platt's scaling~\cite{platt}.
Calibration is also related to generalisation:
Carrell \etal \cite{carrell2022calibration} have shown that the calibration error is bounded by the generalisation error.
This means that better calibrated classifiers are obtained by improving their generalisation.
Calibrated probabilities help with related downstream tasks \cite{bhatt2021uncertainty}.
Here, we focus on reliability estimation, which is useful for rejecting examples for which the model is unsure.
Given the estimated probability $\hat y$ of the audio being fake,
we obtain uncertainty estimates
by computing the entropy: $\hat y \log \hat y + (1 - \hat y) \log (1 - \hat y)$.
We use a unit-scaled variant of the entropy
by dividing it
by the maximum entropy $H_{\max}$, which is obtained for the uniform distribution, $y = 0.5$.

\section{Experimental setup}
\label{sec:setup}

\mypar{Datasets.}
\asvspoof (ASV)~\cite{WANG2020101114} is a popular deepfake detecion dataset,
which we use for training in all our experiments.
The fake audio samples are synthesised with 19 systems
(6 systems in the \texttt{train} and \texttt{dev} splits,
13 in the \texttt{test} split).
The \texttt{train} and \texttt{dev} splits have 50k audio samples (5k real, 45k fake),
while \texttt{test} has 70k samples (63k fake and 7k real).
Similar to~\cite{tak2021icassp}, we use both \texttt{train} and \texttt{dev} for training.

To test the out-of-domain generalisation capabilities we ensemble a benchmark of eight datasets,
including partially-spoofed and multilingual datasets.
Seven of these datasets are summarised in Table~\ref{tab:dataset}.
Additionally, we use the \texttt{aug-dtw} version of TIM (denoted by TIM$^*$ hereafter),
which has identical characteristics to the original dataset, 
except the audios underwent extra post-processing steps,
such as time and pitch shifting, compression, filtering, and dynamic time warping.

\mypar{Metrics.} We evaluate the methods from two perspectives:
(i) their discriminative power over fake and real samples; and
(ii) their ability to produce calibrated predictions.
These desiderata are measured by the \textit{equal error rate} (EER) and \textit{expected calibration error} (ECE)~\cite{naeini2015obtaining}, respectively.
In particular, ECE measures the absolute difference between the average predicted probability and the actual observed frequency within bins of predicted probabilities (we use 15 equally-spaced bins).
The lower the ECE, the better the calibration.

\mypar{Implementation details.}
We extract the self-supervised representations as the average of the last layer of hidden states.
For classification, we use the \texttt{scikit-learn} \cite{scikit-learn} implementation of logistic regression
with $C = 10^6$, implying a low regularisation coefficient,
and set the maximum number of iterations to 1,000.
Training on the full \asvspoof dataset (50k samples) using 1920-dimensional features takes 40 seconds on a 64-core machine.
Our code is available at: \small\url{https://github.com/danoneata/aletheia}.

\section{Experimental results}
\label{sec:results}

\subsection{How well do self-supervised representations generalise?}
\label{subec:sota}

\input{tab-generalisation}

The most important quality of deepfake detectors is their generalisation capability. We measure this by training on the \asvspoof dataset and
evaluating on the benchmark described in Section~\ref{sec:setup}.
The results, which are shown in Table~\ref{tab:sota},
contrast the performance of pretrained self-supervised representations (row 3)
to that of the best models in the literature (row 1)
and to the RawNet2 model (row 2),
one of the best models on generalisation according to M\"uller \etal \cite{muller2022interspeech}.
As self-supervised representation we use the 2B XLS-R variant from wav2vec 2.0, \texttt{wav2vec2/xls-r-2b},
which is the largest model and trained on most data;
we analyze more variants in Section~\ref{susbec:representations}. 

We observe that pretrained representations perform on average much better than RawNet2: 8.8\% EER versus 30.9\% EER.
The performance is also better on each individual dataset with a single exception, TIM.
Our approach also compares favourably to many of the state-of-the-art methods.
These are much more complex methods, which are evaluated only on a handful of datasets.
For several datasets we are the first to either report results (OSDD) or the first to report results in terms of EER (TIM, TIM$^*$, MLAAD).

In terms of computational requirements,
the cost of self-supervised representations is dominated by the feature extraction step.
The time to process an audio of 3 seconds on a Tesla T4 GPU is around 0.3 s with a video memory consumption of around 9GB.
These requirements are around an order of magnitude larger than those of RawNet2,
but still reasonable in the absolute and attainable by commodity hardware.

\subsection{How reliable are the self-supervised representations?}
\label{susbec:reliability-estimation}

\begin{figure}
    \centering
    \includegraphics[trim={5pt 10pt 10pt 5pt}, clip, width=\columnwidth]{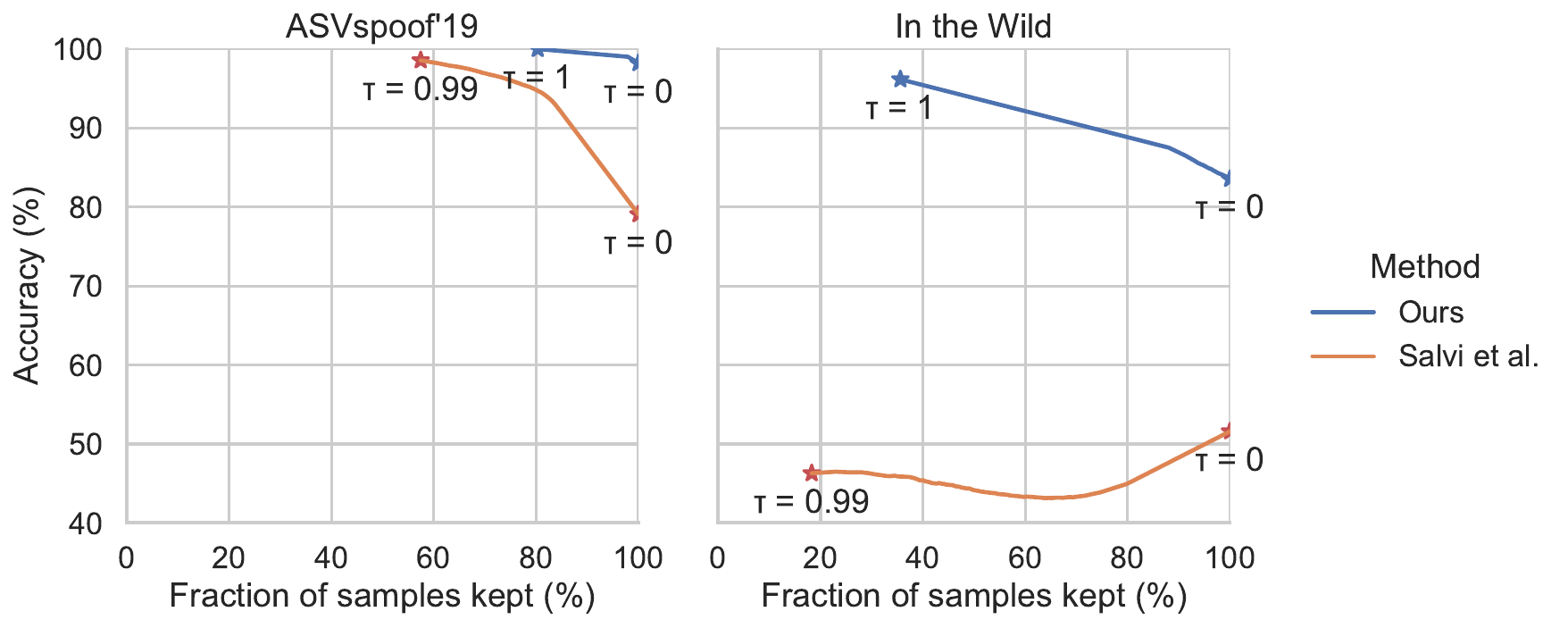}
    \caption{%
        Evaluation of reliability estimation in terms of accuracy and fraction of samples kept,
        as we vary the reliability threshold $\tau \in [0, 1]$.
        Our results (blue) are more reliable than those of Salvi \etal ~\cite{salvi2023icassp} (orange) on both metrics and datasets.
    }
    \label{fig:reliability-estimation}
\end{figure}

Another important characteristic of deepfake detectors is whether we can trust their predictions.
Following Salvi~\etal~\cite{salvi2023icassp}, we formulate this desideratum as the task of \textit{reliability estimation}: 
we want the model to be able to assess the level of confidence in its predictions,
a high confidence indicating a high probability that the prediction is correct.
To this end, we encode the confidence in a prediction using the entropy of the generated probabilities (see Section~\ref{sec:meth}):
if the entropy is close to zero, the model deems the prediction to be reliable;
conversely, high entropy indicates uncertain inputs.
Salvi~\etal~\cite{salvi2023icassp} used a separate network to estimate the reliability which was trained on
features from a pretrained deepfake audio model (in their case, RawNet2) to predict
whether the predictions of the deepfake model are correct or not.

We use two metrics to measure the reliability estimation capabilities \cite{nadeem2009accuracy}:
the fraction of data that is reliable and the classifier accuracy on this data.
For our approach we produce curves by varying a threshold $\tau$ from 0 to 1 in steps of 0.01 on the unit-scaled entropy of each prediction.
For the method of \cite{salvi2023icassp} we vary the threshold similarly, but on the maximum reliability score across segments.
A threshold of 0.5 corresponds to their original evaluation:
if all the audio's segments are unreliable (reliability less than 0.5) then the entire audio is deemed unreliable.

We report results on the \asvspoof and \inthewild datasets,
but differently from \cite{salvi2023icassp}, we evaluate on \textit{all} the samples, not only fake samples.
This is a more fair evaluation, since otherwise a classifier that predicts only fake labels with high confidence will obtain a perfect score.
The results for \cite{salvi2023icassp} are computed over the predictions which were provided by the authors.
The results are shown in Figure~\ref{fig:reliability-estimation}.
We observe much better results than prior work on both datasets, in terms of both metrics, and at all thresholds.
Naturally, there is a drop in performance when going out-of-domain (on \inthewild), but it is less severe than what we observe for the method of Salvi \etal \cite{salvi2023icassp}. 
Moreover, on \inthewild we see that we can trade off data kept for accuracy, which is not the case for the other method.

\subsection{How do other self-supervised representations perform?}
\label{susbec:representations}

\begin{table*}
\newcommand{\ii}[1]{{\scriptsize \color{gray} #1}}
\newcommand{\key}[1]{{\footnotesize \texttt{#1}}}
\centering
\setlength{\tabcolsep}{3pt}
\footnotesize
\begin{tabular}{rlrrclrrrrrrrrrr}
\toprule
&
& Model
& \multicolumn{3}{c}{Pretraining data}
& \multicolumn{5}{c}{EER (\%) ↓} 
& \multicolumn{5}{c}{ECE (\%) ↓} \\
\cmidrule(lr){3-3}
\cmidrule(lr){4-6}
\cmidrule(lr){7-11}
\cmidrule(lr){12-16}
& Name & Size & Dur. (h) & Langs. & Datasets & \multicolumn{1}{c}{ASV} & \multicolumn{1}{c}{ITW} & \multicolumn{1}{c}{TIM} & \multicolumn{1}{c}{FoR} & \multicolumn{1}{c}{Mean}& \multicolumn{1}{c}{ASV} & \multicolumn{1}{c}{ITW} & \multicolumn{1}{c}{TIM} & \multicolumn{1}{c}{FoR} & \multicolumn{1}{c}{Mean}\\
\midrule
\multicolumn{2}{l}{\key{wav2vec2/}} \\
\ii{1}  & \key{base}          &  94M &   1k & en   & LS                       &     4.0 &    45.9 &    59.4 & 36.2 &  36.3 &   5.1 &     61.3 & 6.9 & 38.5 & 27.9\\
\ii{2}  & \key{large}         & 300M &   1k & en   & LS                       &     3.7 &    29.0 &    51.7 & 23.5 &  26.9  &    3.3 &     40.9 & 4.7 & 29.8 & 19.6\\
\ii{3}  & \key{large-lv60}    & 300M &  53k & en   & LL                       &     2.5 &    50.3 &    90.3 & 28.6 &  42.9  &    2.2 &     61.1 & 2.77 & 35.6 & 25.4\\
\ii{4}  & \key{large-robust}  & 300M &  65k & en   & LS, LL, SF, CV/en, VP/en &     2.9 &    21.3 &    52.3 & 12.3 &   22.2  &    4.1 &     31.8 & 16.2 & 9.4 & 15.3\\
\ii{5}  & \key{large-xlsr-53} & 300M &  56k & many & CV, BBL, MLS             &     1.1 &    25.0 & \bf 4.7 & 15.8 &  11.6   &      2.3 &     36.7 & \bf 1.6 & 13.3 & 13.4\\
\ii{6}  & \key{xls-r-300m}    & 300M & 436k & many & CV, BBL, MLS, VP, VL     &     1.0 &    21.3 &    37.8 & \bf 5.1 &  16.2   &       2.2 &     38.7 & 8.4 & 8.4 & 14.4\\
\ii{7}  & \key{xls-r-1b}      &   1B & 436k & many & CV, BBL, MLS, VP, VL     &     1.3 &    18.7 &    13.2 & 9.0 &   10.5  &       3.1 &     26.9 & 8.4 & 11.9& 12.5\\
\ii{8}  & \key{xls-r-2b}      &   2B & 436k & many & CV, BBL, MLS, VP, VL     & \bf 0.6 & \bf 7.2 &    11.3 & 6.8 &   \bf6.5  & \bf 1.8 & \bf 16.1 & 2.3 & \bf 6.3 & \bf 6.6\\
\midrule
\multicolumn{2}{l}{\key{wavlm/}} \\
\ii{9}  & \key{base}             &  94M &  1k & en & LS                          &     4.8 &    37.6 & 50.2 & 44.6& 34.3  &      4.5 &     50.4 & 2.2 & 46.6 & 25.9\\
\ii{10} & \key{base-plus}        &  94M & 84k & en & LS, GS, VP                  &     3.6 &    38.6 & 50.3 & 38.7&   32.8 &    3.0 &     54.6 & 2.3 & 41.8& 25.4\\
\ii{11} & \key{large}            & 300M & 84k & en & LS, GS, VP                  &     1.8 &    32.8 & 55.6 & 8.6 &   24.7 &    1.5 &     55.8 & 2.9 & 23.8 & 21.0\\
\bottomrule
\end{tabular}
\caption{%
    Evaluation of self-supervised representations for deepfake detection.
    By keeping various dimensions fixed,
    we can asses the impact of
    model size (rows 1--2, 6--8, 10--11),
    pretraining data (rows 2--6, 9--10),
    and representation type (row 1 vs. row 9).
    The pretraining datasets are
    LibriSpeech (LS),
    LibriLight (LL),
    Switchboard \& Fisher (SF),
    CommonVoice (CV),
    VoxPopuli (VP),
    Babel (BBL),
    Multilingual LibriSpeech (MLS),
    VoxLingua107 (VL),
    GigaSpeech (GS). 
}
\label{tab:representations}
\end{table*}

\begin{figure}
    \centering
    \includegraphics[width=\columnwidth]{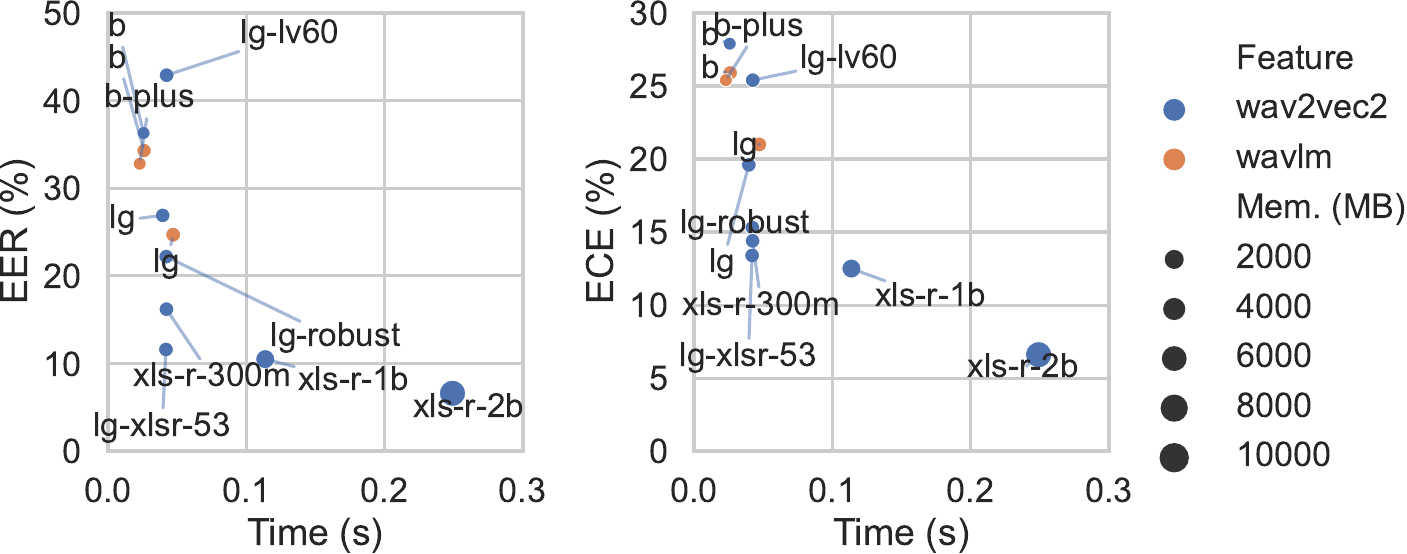}
    \caption{%
    Performance trade off as a function of inference time for the 11 variants of self-supervised representations.
    Marker area indicates peak video memory.
    Time and memory are averaged over 64 audio files, which average three seconds.}
    \label{fig:feature-type-benchmark}
\end{figure}

Pretrained self-supervised representations come in multiple variants, differing in terms of architecture, model size or pretraining data.
Based on this information we decouple the performance on three axes: model family, model size, data.
Table~\ref{tab:representations} shows the results for 11 variations of self-supervised representations belonging to two classses of models: \wavvec and \wavlm.
The results are given in terms of discrimination (EER) and calibration error (EER) on 4 out of the 8 datasets previously used.

We observe that both performance metrics improve on average with the size of the self-supervised model.
When increasing the model size from 94M to 300M parameters,
the mean EER decreases from 36.3\% to 26.9\% (rows 1--2) for \wavvec and from 32.8\% to 24.7\% (rows 10--11) for \wavlm.
The error further decreases from 16.2\% to 6.6\% (rows 6--8) when increasing the \wavvec model size from 300M to 2B parameters.
The best mean ECE is also obtained for the largest 2B model: 6.6\%.

Among the two families of representations, \wavvec appears to be the better representation for deepfake detection.
While \wavlm is only slightly better than \wavvec at 94M parameters (34.3\% EER, row 9, vs. 36.3\% EER, row~1),
\wavvec has a much improved 300M-parameter model: 11.6\% EER (row 5) vs. 24.7\% EER (row 11).
Similar observations can be made for the calibration performance.

The conclusions regarding the pretraining data are not as clear.
Sometimes more diverse training data improves performance (row 4 vs. rows 2--3; row 10 vs row 9), but in other cases adding more data just hurts performance (row 6 vs. row 5).
We find this result surprising and worth of further investigation.

In Figure~\ref{fig:feature-type-benchmark}, we also analyze the computational requirements of these representations.
We observe that the best performing representation, \texttt{xls-r-2b}, has the largest costs in terms of both memory and time, albeit still reasonable in the absolute.
In order to optimise different aspects, one may choose other representations, for example one with a good trade-off among the considered metrics is the \texttt{large-xlsr-53} variant.

\subsection{How important is the classifier?}

We have experimented with two more flexible models on top of the frozen \texttt{wav2vec2/xls-r-2b} representations:
a three-layer multilayer perceptron (MLP) with ReLU activation, and
a self-attention layer (SAL) followed by a linear layer.
Additionally, we have investigated a stronger regularisation value for logistic regression, $C = $ 1.
Results are shown in Table~\ref{tbl:classifiers}.
On average, logistic regression obtains best results on both metrics, although there are variations across the test datasets.
Surprisingly, the less regularised variant, $C = 10^6$, generalises better.
The reason is perhaps that the logistic model, being a linear model, is already highly constrained,
so further regularisation ($C = 10^0$) results in underfitting and poorer results.

\begin{table}[htb]
    \newcommand{\ii}[1]{{\tiny \color{gray} #1}}
    \scriptsize
    \centering
    \setlength{\tabcolsep}{2.3pt}
    \begin{tabular}{lr rrrrr rrrrr}
    \toprule
     & & \multicolumn{5}{c}{EER (\%) ↓} & \multicolumn{5}{c}{ECE (\%) ↓} \\
    \cmidrule(lr){3-7}
    \cmidrule(lr){8-12}
     Classifier & $C$
     & \multicolumn{1}{c}{ASV} & \multicolumn{1}{c}{ITW} & \multicolumn{1}{c}{TIM} & \multicolumn{1}{c}{FoR} & \multicolumn{1}{c}{$\mu$}
     & \multicolumn{1}{c}{ASV} & \multicolumn{1}{c}{ITW} & \multicolumn{1}{c}{TIM} & \multicolumn{1}{c}{FoR} & \multicolumn{1}{c}{$\mu$}
     \\
     \midrule

    SelfAtt & & 1.3          & 9.3  & 26.59  & 8.1  & 11.3  
                   & 1.5          & \textbf{7.0} & 33.2 & 18.7 & 15.1
                   \\
    MLP    &       & \textbf{0.4} & 10.3 & 17.2 & \textbf{4.9} & 8.2
                   & \textbf{0.8} & 20.7 & 4.2 & 8.0 & 8.4
                   \\
    LogReg & $10^0$ & \textbf{0.4}          & 8.6  & 19.1 & 7.8  & 9.0 
                    & 1.4          & 14.5 & \textbf{2.3} & \textbf{6.3}  & \textbf{6.1}
                    \\ 
    LogReg & $10^6$ & 0.6          & \textbf{7.2} & \textbf{11.3} & 6.8  & \textbf{6.5}
                    & 1.8          & 16.1 & \textbf{2.3} & \textbf{6.3} & 6.6
                          \\
    \bottomrule
    \end{tabular}
\caption{
Performance of different classifiers trained on the \texttt{wav2vec2/xls-r-2b} representation. %
All classifiers are trained on the same data: ASVSpoof'19 \texttt{train} and \texttt{dev} splits.
}
\label{tbl:classifiers}
\end{table}

\subsection{What is the impact of training data?}
\label{susbec:training-data}

\begin{figure}[htb]
    \centering
    \includegraphics[trim={15pt 15pt 15pt 15pt}, clip, width=\columnwidth]{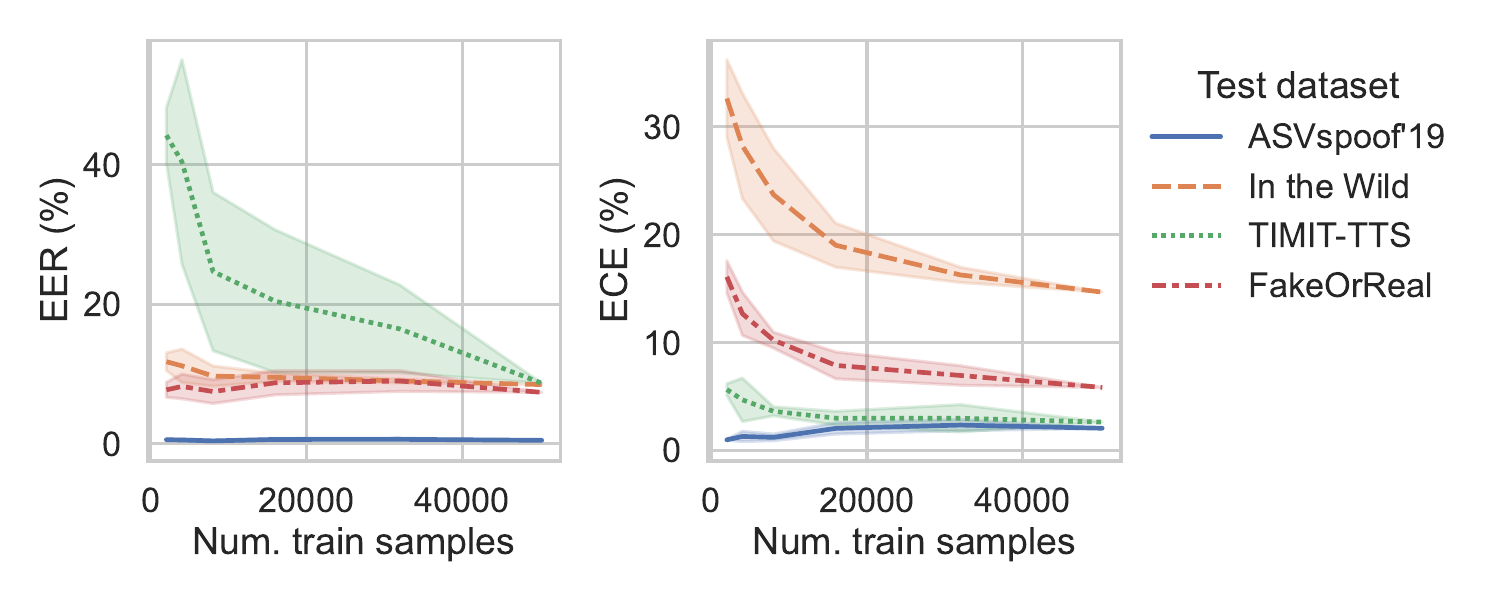}
    \caption{%
        Performance versus number of \asvspoof training samples
        using \texttt{wav2vec2/xls-r-2b}.
        Error bars are one standard deviation over three random subsets of training data.
    }
    \label{fig:num-samples}
\end{figure}

A benefit of using a simple linear layer classifier is that we can expect the model to be robust to the number of training samples.
We test this hypothesis by varying the number of training samples from \asvspoof in factors of two, \{2k, 4k, 8k, 16k, 32k, 50k\}.
In Figure \ref{fig:num-samples} we show that the plots exhibit a kink at around 8k indicating that even with significantly fewer samples than all 50k,
we can attain good performance on out-domain data.

\section{Conclusions}
\label{sec:conc}

We investigated self-supervised representations for audio deepfake detection 
in terms of two important properties:
generalisation and calibration.
The employed approach---self-supervised representations, followed by a linear layer---attained state-of-the-art results
on a benchmark of eight out-of-domain datasets, as well as on the task of reliability estimation.
In terms of the type of representations, we have showed the importance of features from large models (the 2B-parameter variant of wav2vec 2.0).
In terms of the classifier, a linear layer ensured strong out-of-domain performance provided it was not highly regularised.

\mypar{Limitations.}
Despite the strong results, our study leaves an important question open:
Why do we sometimes see inconsistent results across datasets?
This lack of understanding is not specific to our work,
but underpins deepfake detection methods in general \cite{cuccovillo2022challenges, yi2023survey}.
Therefore, a worthwhile future endeavour is to provide a better understanding of what deepfake classifiers learn.

\section{Acknowledgements}
This work was funded by EU Horizon projects AI4TRUST (No. 101070190) and ELIAS (No. 101120237), and by CNCS/CCCDI UEFISCDI (No. PN-IV-P8-8.1-PRE-HE-ORG-2023-0078).

\bibliographystyle{IEEEtran}
\bibliography{refs}

\end{document}

%% file: tab-datasets.tex
\begin{table*}
    \footnotesize
    \newcommand{\na}{{\color{gray}\textsc{n/a}}}
    \newcommand{\ii}[1]{{\scriptsize \color{gray} #1}}
    \newcommand{\lbl}[1]{{\scriptsize \color{gray} #1}}
    \newcommand{\key}[1]{{\footnotesize \texttt{#1}}}
    \newcommand{\dur}[2]{#1{\scriptsize ±#2}}
    \centering
    \begin{tabularx}{\textwidth}{lXccrrrrrr}
\toprule
Dataset                          & Real data                                      & Langs.     & Partial & Systems      & SR        & Real        & Fake        & Duration \\
\lbl{short name}                 &                                                &            &         &              & \lbl{kHz} & \lbl{count} & \lbl{count} & \lbl{seconds} \\
\midrule
ASV~\cite{WANG2020101114}        & VCTK                                           & en         & \xmark  & 10 TTS, 3 VC & 16        & 7k          & 63k         & \dur{3.1}{2.9} \\
ITW~\cite{muller2022interspeech} & YouTube                                        & en         & \xmark  & \na          & 16        & 20k         & 12k         & \dur{4.2}{6.6} \\
TIM~\cite{timittts}              & VidTIMIT                                       & en         & \xmark  & 12 TTS       & 16        & 430         & 20k         & \dur{3.1}{2.3} \\
FoR~\cite{for}                   & Arctic, LJSpeech, VoxForge, YouTube, TED Talks & en         & \xmark  & 6 TTS        & 16        & 34k         & 34k         & \dur{3.0}{4.5} \\
PS~\cite{parspoof}               & VCTK                                           & en         & \cmark  & 19 TTS or VC & 16        & 7k          & 63k         & \dur{3.4}{3.5} \\
ODSS~\cite{odss}                 & VCTK, Hi-Fi TTS, HUI-ACG, SLR-ES               & en, es, de & \xmark  & 2 TTS        & 16        & 11k         & 19k         & \dur{3.1}{4.1} \\
MLAAD~\cite{mlaad}               & M-AILABS                                       & many (23)  & \xmark  & 52 TTS       & 22        & 20k         & 80k         & \dur{7.6}{10.1} \\
\bottomrule
    \end{tabularx}
    \caption{%
        Datasets from the evaluation benchmark used in this paper.
        We consider only the test splits of these datasets.
        The duration of the audio files is given as the average plus-minus two times standard deviation. 
    }
    \label{tab:dataset}
\end{table*}

%% file: tab-generalisation.tex
\begin{table*}
    \footnotesize
    \setlength{\tabcolsep}{3.5pt}
    \newcommand{\na}{{\color{gray}\textsc{n/a}}}
    \newcommand{\ii}[1]{{\scriptsize \color{gray} #1}}
    \newcommand{\key}[1]{{\footnotesize \texttt{#1}}}
    \centering
    \begin{tabular}{rlrrrr
    r@{\hspace{0.0\tabcolsep}}l 
    r@{\hspace{0.0\tabcolsep}}l 
    r@{\hspace{0.0\tabcolsep}}l 
    r@{\hspace{0.0\tabcolsep}}l 
    r@{\hspace{0.0\tabcolsep}}l 
    r@{\hspace{0.0\tabcolsep}}l 
    r@{\hspace{0.0\tabcolsep}}l 
    r@{\hspace{0.0\tabcolsep}}l 
    r}
        \toprule
        &
        & \multicolumn{1}{c}{Params.} 
        & \multicolumn{2}{c}{Time}
        & \multicolumn{1}{c}{Memory}
        & \multicolumn{17}{c}{EER (\%) $\downarrow$} \\
        \cmidrule(lr){4-5}
        \cmidrule(lr){6-6}
        \cmidrule(lr){7-23}
        &
        Method
        & \multicolumn{1}{c}{\scriptsize \it trainable}
        & \multicolumn{1}{c}{train}
        & \multicolumn{1}{c}{pred}
        & \multicolumn{1}{c}{pred}
        & \multicolumn{2}{c}{ASV}
        & \multicolumn{2}{c}{ITW}
        & \multicolumn{2}{c}{TIM}
        & \multicolumn{2}{c}{TIM$^*$}
        & \multicolumn{2}{c}{FoR}
        & \multicolumn{2}{c}{PS}
        & \multicolumn{2}{c}{OSDD}
        & \multicolumn{2}{c}{MLAAD}
        & Mean \\
        \midrule
        \ii{1} & State of the art & & & & 
        & \bf 0.2 & $\,$\cite{Nautsch_2021}
        & 7.7 & $\,$\cite{lu2023oneclass}
        & \na & 
        & \na & 
        & 18.1 & $\,$\cite{zhang2021fake} 
        & 14.2 & $\,$\cite{parspoof}
        & \na  &
        & \na \\
        \midrule
        \ii{2} & RawNet2                                  & 25M & 8h & 0.03s & 1.3GB & 5.9 &  \scriptsize±0.1   & 46.7 &  \scriptsize±0.3    & \bf 2.4  &  \scriptsize±0.3    & 27.9 &  \scriptsize±0.5     & 52.1 & \scriptsize ±0.2     & 33.1  &  \scriptsize±0.3    & 45.0 & \scriptsize ±0.3     & 34.4 &  \scriptsize±0.2    & 30.9 \\
        \ii{3} & Ours                  & 2k & 4h & 0.26s & 9.3GB & 0.5 & \scriptsize ±0.1 & \bf 7.2  & \scriptsize ±0.3 & 11.5 & \scriptsize ±1.1 & \bf 3.6  & \scriptsize ±1.4 & \bf 6.9  & \scriptsize ±0.2 & \bf 5.1  & \scriptsize ±0.9 & \bf 16.0 & \scriptsize ±0.4 & \bf 20.0 & \scriptsize ±0.3 & \bf 8.8 \\
        \bottomrule
    \end{tabular}
    \caption{%
    Comparison in terms of EER with state-of-the-art on multiple out-of-domain datasets.
    We report mean and two times standard deviation using ten runs of bootstrap on the test data. On average, the proposed approach improves from $30.9\%$ EER to $8.8\%$ EER compared to RawNet2. We also show significant improvements when compared to available state-of-the art methods for each dataset.
    }
    \label{tab:sota}
\end{table*}